\documentclass[showpacs,aps,prd,preprintnumbers,twocolumn]{revtex4-1}
\usepackage{epsfig}
\usepackage{url}
\usepackage{array,tabularx,epsfig,mathrsfs,graphicx,rotating}
\usepackage{ifthen}
\usepackage{fancybox}
\usepackage{amsfonts}
\usepackage{url}
\usepackage{hyperref}
\hypersetup{
  colorlinks=true,
  linkcolor=blue,
  citecolor=blue,
  urlcolor=blue
}
\newcommand{\beq}{\begin{equation}}
\newcommand{\eeq}{\end{equation}}

\newcommand{\ptjet}{p_{T}(\mathrm{jet})}
\newcommand{\pttop}{p_{T}(\mathrm{top})}

\newcommand{\etajet}{\eta(\mathrm{jet})}

\newcommand{\mtop}{m(\mathrm{top})}
\newcommand{\mjet}{M_{\mathrm{jet}}}
\chardef\til=126

\newcommand{\pythia}{{\tt PYTHIA8\,}}
\newcommand{\herwig}{{\tt HERWIG++\,}}
\usepackage{subfigure}

\date{\today}

\begin{document}

\preprint{ANL-HEP-PR-13-5}

\title{Studies of  highly-boosted top quarks  near the TeV scale  using \\ jet masses
at the LHC}

\author{B.~Auerbach, S.V.~Chekanov}
\affiliation{ 
HEP Division, Argonne National Laboratory,
9700 S. Cass Avenue, 
Argonne, IL 60439, 
USA
}

\email[E-mail:]{ben.auerbach@gmail.com}

\email[E-mail:]{chekanov@anl.gov}

\author{N.~Kidonakis}
\affiliation{
Kennesaw State University, Physics \#1202,
1000 Chastain Rd., Kennesaw, GA 30144-5591, USA
}

\email[E-mail:]{nkidonak@kennesaw.edu}

\begin{abstract}
Studies of highly-boosted top quarks produced inclusively in $pp$ collisions
at $\sqrt{s}=14$~TeV are discussed. 
Using Monte Carlo models
after a fast detector simulation,
it is shown that jet masses alone provide a sensitive probe for top quarks produced 
inside high-$p_T$ jets. The hadronic decays of such top quarks can be studied in a data-driven approach
by analysing shapes of jet-mass distributions. 
It is shown that inclusive production of boosted top quarks can be
observed if it has a cross section at least twice larger than 
the prediction from the approximate next-to-next-to-leading-order (aNNLO) calculation
for the  $t\bar{t}$ process. The $t\bar{t}$ process 
with the nominal aNNLO strength can be measured using the masses of jets after a $b$-tagging.  
\end{abstract}



\maketitle

%%%%%%%%%%%%%%%%%%%%%%%%%%%%%%%%%%%%%%%%%%%%%%%%%%%%%%%%%%%%%%%%%%
\section{Introduction}
%%%%%%%%%%%%%%%%%%%%%%%%%%%%%%%%%%%%%%%%%%%%%%%%%%%%%%%%%%%%%%%%%%

Heavy particles with masses above a TeV decaying to top quarks
can lead to
an enhanced cross section of top quarks  compared to the Standard Model expectations.
The fact that such a cross section can be more than doubled  
for top quarks with high transverse momenta ($\pttop$) was recognized \cite{PhysRevD.49.4454} 
almost immediately after the discovery of top quarks at the Tevatron.
However, the Standard Model predictions on top quark cross section 
have not yet been confronted with experimental data 
for transverse energies close to the TeV scale.

According to the Standard Model, inclusive 
production of top quarks is dominated by the $t\bar{t}$ process. 
Top-quark production   
includes contributions
from single top quark processes ($t$- and $s$-channels) and from  $Wt$.
Top quarks can also be produced via associated Higgs production. Finally,
top quarks at very large $\ptjet$ can originate from fragmentation, but
no data exist to constrain this process.

Currently, there are several high-$p_T$ measurements of top quarks focusing on the $t\bar{t}$
event topology. 
The D0 collaboration has reported the $t\bar{t}$  cross
section up to  $\pttop=350$~GeV \cite{Abazov:2010js}.
The CDF collaboration \cite{cdftop} performed searches  for
highly-boosted top quarks, but statistics was insufficient to support
the claim for observation of top-quark production at $\pttop>400$ GeV.
At the LHC, ATLAS performed \cite{:2012qa} searches for $Z'$ extending the reach in $\pttop$ up to
$500$~GeV, but without cross section measurements. 
CMS recently measured the top quark $p_T$ distribution up to $\pttop=400$~GeV \cite{:2012qka}. 

The measurement of top-quark cross sections at very large transverse momenta is challenging.
For large jet transverse momenta, the identification
of leptons (muons and electrons) from the $W$ decay is difficult
since they are often collimated with
$b-$jets from the top decays.
This leads to a reduced electron efficiency due to isolation requirements and
large fake rates for muons due to the presence of $b$-quark decay products.
In addition, a $b$-tagging technique 
suffers from an inefficiency for large $\ptjet$ and poor separation between the signal and
multijet background events.

Because of the above reasons,
the main focus of this analysis is the 
hadronic-final state characteristics of jets which are expected to be
sensitive to the production of hadronically decaying top quarks with large $\pttop$.
For such studies, jet masses and jet shapes are often discussed as a useful tool
for the identification of top quarks and  
for reduction of the overwhelming rate from conventional QCD processes 
\cite{Agashe:2006hk,*Lillie:2007yh,*Butterworth:2007ke,*Almeida:2008tp,*Almeida:2008yp,
*Kaplan:2008ie,*Brooijmans:2008,*Butterworth:2009qa,*Ellis:2009su,*ATL-PHYS-PUB-2009-081,*CMS-PAS-JME-09-001,*Almeida:2010pa,*Hackstein:2010wk,Chekanov:2010vc,*Chekanov:2010gv}.

In this paper, we adopt a strategy 
based on a high-precision measurement of shapes of jet masses. 
Using realistic Monte Carlo (MC) simulations after a fast detector simulation, 
we show that hadronic decays of highly-boosted top quarks  
can be observed by performing a data-driven analysis of jet-mass shapes near the 170 GeV region,
without any additional technique involving 
jet substructure variables.
This article shows that this method becomes feasible 
if the top-quark yield in the fiducial region  $\pttop>0.8$~TeV 
is a factor two or more larger than the expectation from the best understood 
$t\bar{t}$ process. Given large theoretical uncertainties for
the $t\bar{t}$ process at large $\pttop$ and  
a number of other not well understood sources (see Sect.~\ref{sec:theory}) contributing to top 
quark production at large $\pttop$, this approach can be promising 
for observation of inclusively produced top quarks. 
Moreover, we also demonstrate that a $b-$tagging can substantially increase 
the signal-over-background ratio, leading to observation of top jets from $t\bar{t}$.

%%%%%%%%%%%%%%%%%%%%%%%%%%%%%%%%%%%%%%%%%%%%%%%%%%%%%%%%
\subsection{Theoretical calculations for inclusive top production}
\label{sec:theory}
%%%%%%%%%%%%%%%%%%%%%%%%%%%%%%%%%%%%%%%%%%%%%%%%%%%%%%%%

\begin{figure}
\begin{center}
\includegraphics[scale=0.38, angle=0]{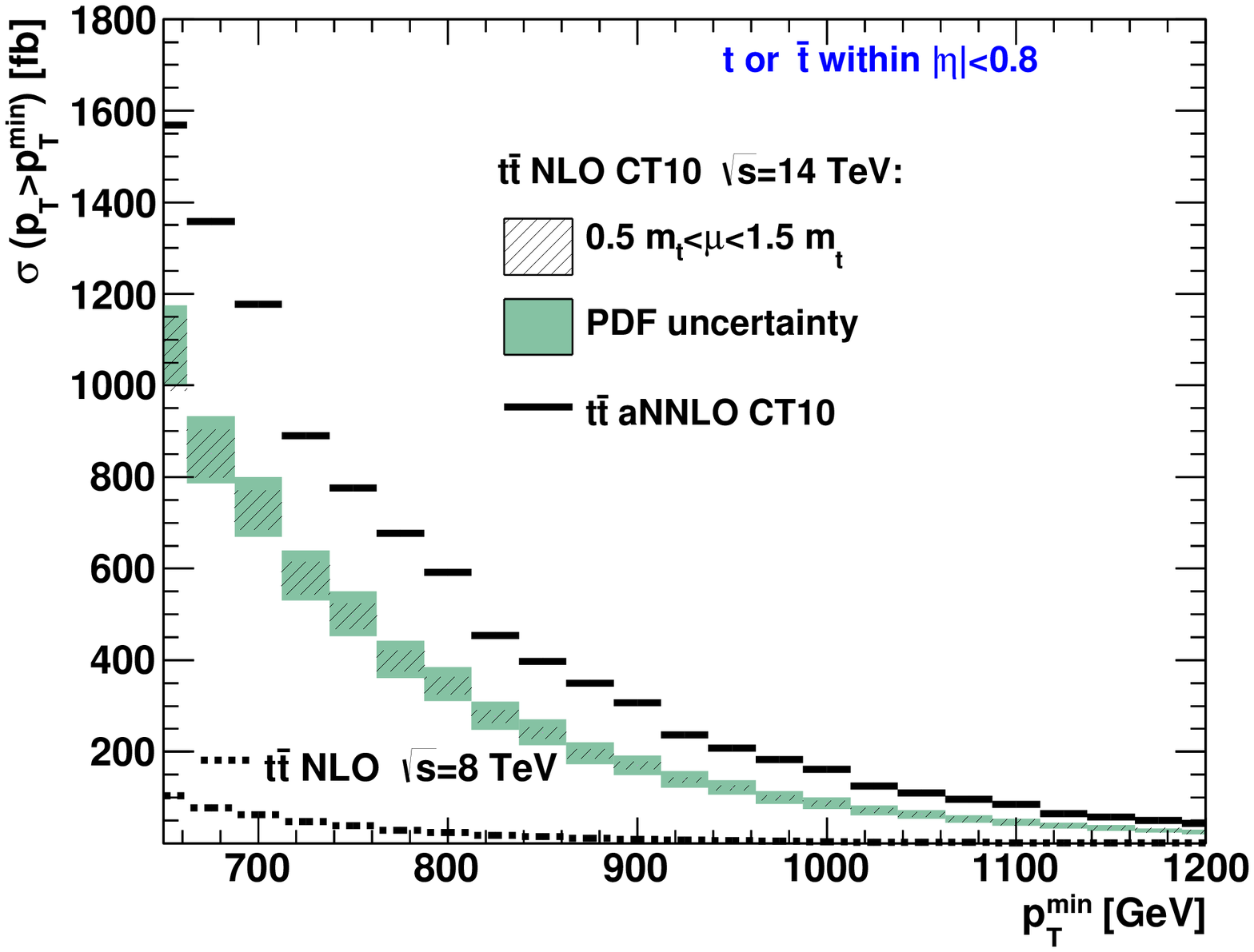}
\end{center}
\caption{
The NLO and aNNLO cross sections 
for the number of top quarks in the $t\bar{t}$ process as a function of transverse momenta
for  $|\eta|<0.8$.
The hatched area shows the renormalization scale uncertainty for NLO, while
the filled green area shows the PDF uncertainty (see the text of Sect.~\ref{sec:theory}).
The dashed line shows the NLO cross section for $\sqrt{s}=8$~TeV.
}
\label{fig:nlo}
\end{figure}

There are several Standard Model processes
contributing to inclusive top-quark production at large $\ptjet$.
The best studied process is the  $t\bar{t}$ process when each boosted top quark
gives origin to a jet.
Single-top production ($t-$ and $s-$channels)
and top-quark associated production are other sources of top-quark jets.
In this case, no  second jet originated from a hadronically decaying  top quark is expected.
Top quarks within a single high-$p_T$ jet can be produced due to flavor-changing processes and 
fragmentation.  
Finally, new resonance physics most readily contributes to the high-$p_T$ region. 

For the present analysis,  the theoretical calculation for high-$p_T$ top quarks
was performed at next-to-leading-order (NLO)  using the  {\tt MCFM} 6.3 program \cite{Campbell201010}
based on the CT10 parton density functions (PDF) \cite{Lai:2010vv}.
The renormalization ($\mu_R$)
and factorization ($\mu_F$) scales were varied between $\mtop-\mtop/2$ and $\mtop+\mtop/2$,
keeping the renormalization and factorization scales to be the same.
The PDF uncertainty was
calculated from 53 CT10 PDF sets.
A check was performed with the {\tt POWHEG} program \cite{nason2007manual} 
which uses a $p_T$ dependent (dynamic) scale
(which is considered to be more appropriate at large $\pttop$).
It  was found that this model is in good agreement with the {\tt MCFM} prediction assuming
the estimated renormalization and factorization scale uncertainties. 

Near partonic threshold for $t{\bar t}$ production the contributions from soft-gluon emission become dominant.
The soft-gluon corrections to the double-differential top cross section in transverse momentum and rapidity can be resummed at next-to-next-to-leading-logarithm
(NNLL) accuracy via the two-loop soft anomalous dimension matrices \cite{Kidonakis:2010dk,*Kidonakis:2012rm}. The resummed result has been expanded at fixed order to
next-to-next-to-leading order (NNLO) and, after integration over rapidity, used to calculate the top quark transverse momentum distribution, $d\sigma/dp_T$. 
This approximate next-to-next-to-leading-order (aNNLO) calculation from NNLL soft-gluon resummation 
leads to a factor of two larger $t\bar{t}$ cross section at large $\pttop$ compared to NLO.

Figure~\ref{fig:nlo} shows the NLO and aNNLO cross sections 
for top quarks from the $t\bar{t}$ process 
in $pp$ collisions at the  center-of-mass energy $\sqrt{s}=14$~TeV.
The cross sections are presented
as a function of the transverse momentum cut in the pseudorapidity region $|\eta|<0.8$.
The expected PDF uncertainty is about $20\%$ (shown as filled band on Fig.~\ref{fig:nlo}), 
while the renormalization scale uncertainty is smaller.
For a comparison, the cross section for $\sqrt{s}=8$~TeV is also shown but without
uncertainties. 
Assuming an integrated luminosity of 10~fb$^{-1}$, the NLO calculation 
predicts 3500 top quarks in the all decay channel in the fiducial volume $\pttop>0.8$~TeV.
This number is expected to increase to 
5920 top quarks for the aNNLO.
The contribution to the top quark yield from single-top production ($t-$channel \cite{Kidonakis:2011wy}, $s-$channel \cite{Kidonakis:2010tc}, and $Wt$ production \cite{Kidonakis:2010ux}) is  expected to be smaller \cite{Kidonakis:2012rm} than for the $t\bar{t}$ process.

Although the main focus of this study are jets with $\ptjet>0.8$~TeV,
it should be pointed out that contributions to such jets from top quarks with $\pttop$ 
lower than $0.8$~TeV are possible due to jet-energy resolution effects.  
In order to take into account such effects,
top quarks were generated at lower $\pttop$ than the minimum $\ptjet=0.8$~TeV used in this analysis. 
In the following studies, top quarks were generated using MC models
with $\pttop>0.65$~TeV. Then their rate was scaled to  15690 top quarks as predicted by the aNNLO  
for the fiducial region $\pttop>0.65$~TeV 
assuming an integrated luminosity of
10~fb$^{-1}$.

\begin{figure}
\begin{center}
\includegraphics[scale=0.38, angle=0]{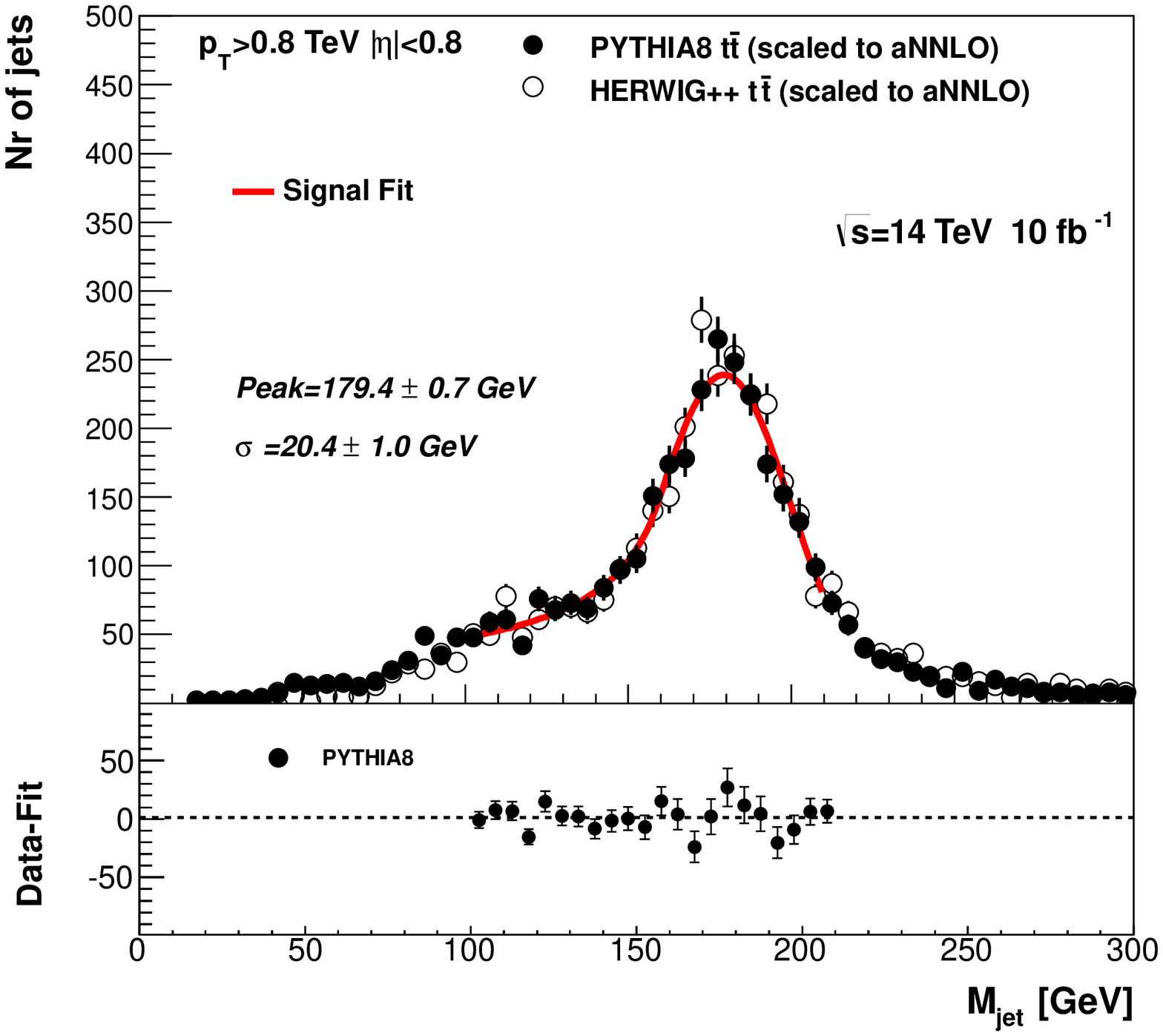}
\end{center}
\caption{
Expectations for the jet mass distribution initiated by top quarks
using \pythia  and \herwig after the fast detector simulation.
The jet selection cuts are $\ptjet>0.8$ TeV and $|\etajet|<0.8$. The 
number of initial top quarks is normalized to the aNNLO for
$\pttop>0.65$~TeV. The expected number of top jets shown in this
figure is 3,500, with 2,200 in the Gaussian core $140<\mjet<200$ GeV. 
The jet masses generated with  \pythia were fitted in the mass range 100-210~GeV using a
Crystal Ball function \protect\cite{Oreglia}.
The bottom plot shows the fit residuals.
The fit has $\chi^2/$ndf=1.3. 
}
\label{fig:tt}
\end{figure}

%%%%%%%%%%%%%%%%%%%%%%%%%%%%%%%%%%%%%
\subsection{Monte Carlo simulations}
%%%%%%%%%%%%%%%%%%%%%%%%%%%%%%%%%%%%%

Top quark jets in  $pp$ collisions
were modeled  using \pythia \cite{Sjostrand:2006za} and \herwig  \cite{Bahr2008}
MC models  assuming $pp$ collisions at a center-of-mass energy of $\sqrt{s}=14$~TeV.
As discussed above, the  number of top quarks in the fiducial region $\pttop>0.65$~TeV 
was scaled to the aNNLO cross section assuming an 
integrated luminosity of 10~fb$^{-1}$.

In addition to the top-quark initiated jets, QCD background due to jets
originating from light-flavor quarks and gluons were considered. 
Hadronic jets from all QCD processes (but excluding the $t\bar{t}$ production),  
were generated using  \pythia and \herwig.
The MC inclusive cross section of jets was  corrected to match the NLO prediction 
estimated with the {\tt NLOjet++} program \cite{Catani:1996vz,*Nagy:2003tz}. 
The estimated scaling factor was found to be close to $10\%$.

The samples for $t\bar{t}$ and for the QCD dijet background
events were processed through  a fast detector simulation based the  {\tt DELPHES} 2.0.3 framework \cite{Ovyn:2009tx}
assuming the ATLAS detector geometry.
The most crucial in such simulation are detector resolutions for 
hadronic and electromagnetic calorimeters of the ATLAS detector. Those were taken from the default
{\tt DELPHES} setting based on the ATLAS studies \cite{Aharrouche:2006nf,Kulchitsky:2000gg}.

%%%%%%%%%%%%%%%%%%%%%%%%%%%%%%%%%%%
\subsection{Jet mass reconstruction}
%%%%%%%%%%%%%%%%%%%%%%%%%%%%%%%%%%%

Events after the fast detector simulation 
were  selected if they contain at least one jet reconstructed
with the anti-$k_T$ algorithm \cite{Cacciari:2008gp}
with  a distance parameter of 0.6.
This distance parameter is the most optimal to 
collect the decay products of hadronically decaying top quarks inside jets with
$\ptjet>0.8$~TeV \cite{Chekanov:2010vc,*Chekanov:2010gv}. 
Jets were reconstructed with the {\tt FastJet}  package \cite{Cacciari:2011ma} using
the {\tt DELPHES} calorimeter cell positions and energies.

The final jets were selected with $\ptjet>0.8$~TeV and 
$|\etajet|<0.8$. 
For the current analysis, the central calorimeter region
is used in order to avoid biases in the reconstruction of jet shapes and
in order to increase the signal over background ratio for
boosted top searches: for $\pttop>0.8$~TeV, top quarks from the $t\bar{t}$ process are
predominantly produced in the very central rapidity region.

%%%%%%%%%%%%%%%%%%%%%%%%%%%%%%%%%%%%%%%%%%
\section{Results}
%%%%%%%%%%%%%%%%%%%%%%%%%%%%%%%%%%%%%%%%%

\subsection{Masses of top jets}

As is well known, jet masses are sensitive to the presence of top-quark decays.
Figure~\ref{fig:tt} shows the masses ($\mjet$)  
of jets initiated by top quarks (``top jets'') 
using \pythia  and \herwig after the fast 
detector simulation. The jet selection cuts are $\ptjet>0.8$~TeV and $|\etajet|<0.8$. 
The jet masses
include contributions from all-top decays (including leptonic decays of $W$ bosons). 
The jet mass distribution can be described by a Crystal Ball function \cite{Oreglia}
which  has a Gaussian core (with a mean $m_0$ and a width $\sigma$) and a
power-law tail with an exponent $n$ to account for energy losses of hadronic decays 
or leptonic $W$ decays.
The parameter $\alpha$ defines the transition between the Gaussian and
the power-law functions.
Figure~\ref{fig:tt} shows the fit with the Crystal Ball function using \pythia. 
The peak position of the Gaussian component,  which is intended to describe fully-hadronic decays,  
is close to 180~GeV with  the width $\sigma \simeq 20$ GeV.
Figure~\ref{fig:tt} shows that the difference in shapes between \pythia and \herwig
is small and thus can be neglected.

\begin{figure}
\begin{center}
\subfigure[Jet mass distribution without $t\bar{t}$ process.]{
\includegraphics[scale=0.36, angle=0]{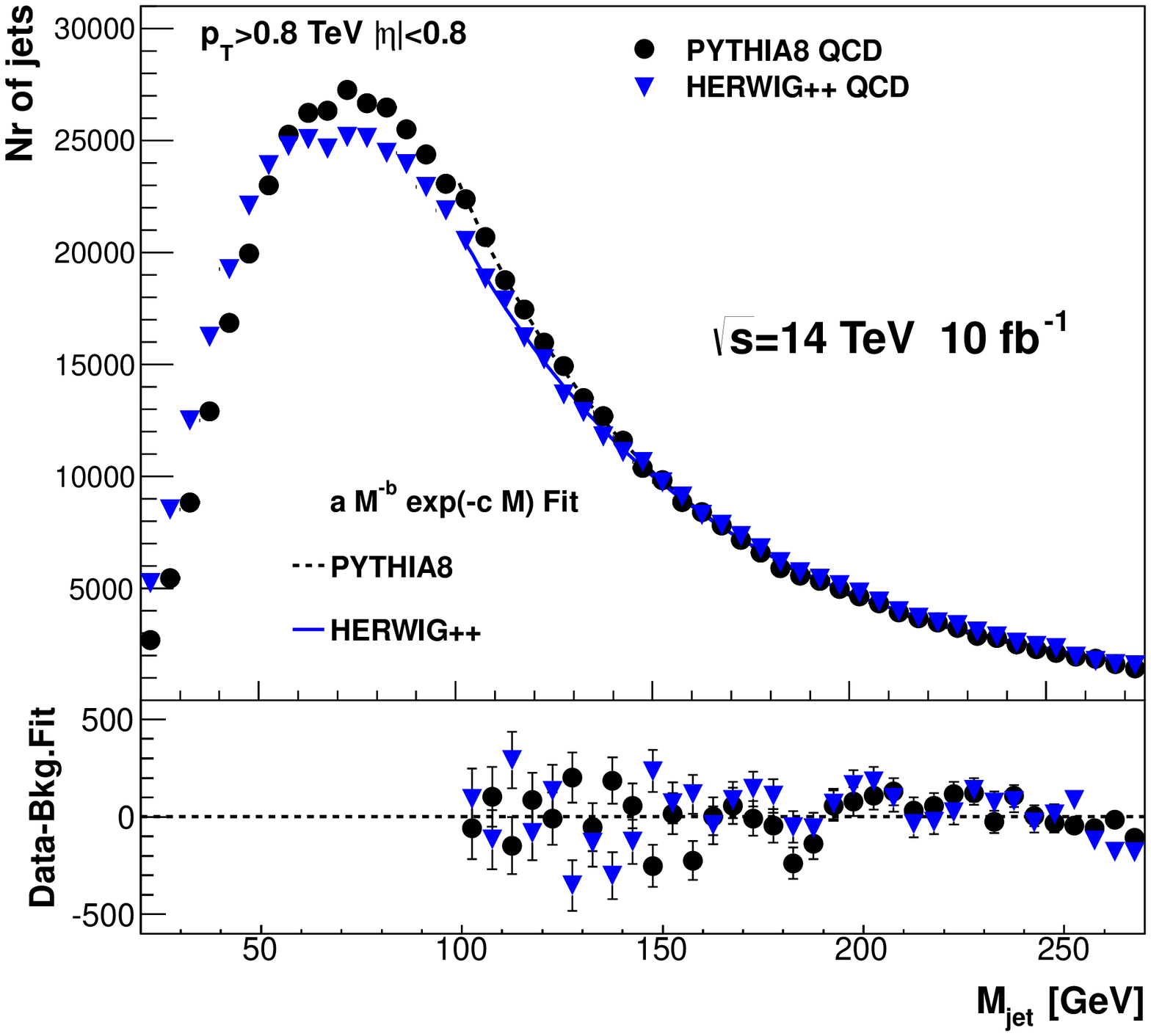}
}
\subfigure[Jet mass distribution with the $t\bar{t}$ process.]{
\includegraphics[scale=0.36, angle=0]{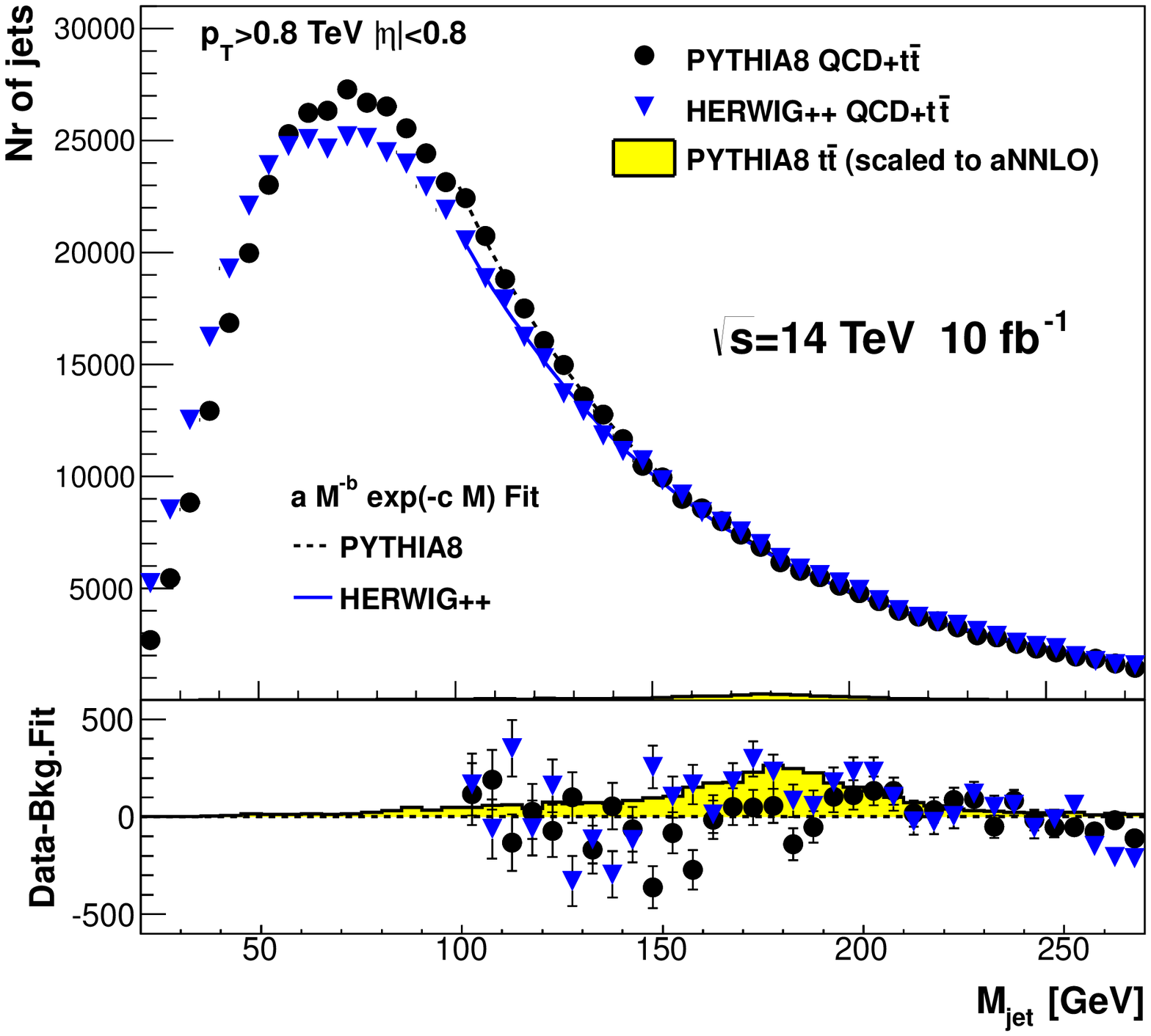}
}
\end{center}
\caption{
Expectations for the jet-mass distributions for the MC models after the fast detector simulation.
The rate of light-flavor jets is scaled to the NLO prediction for
inclusive jets.
The jet masses are shown for (a) assuming no $t\bar{t}$ process and, (b) with the $t\bar{t}$ process included.
A $\chi^2$ fit was performed using the background function $a\cdot \mjet^{-b} \cdot \exp{(-c\cdot\mjet)}$
in the mass range $100<\mjet<270$~GeV. 
The jet mass prediction shown in (b) as shaded histograms
is based on  \pythia $t\bar{t}$ scaled to the aNNLO.
The fit quality is $\chi^2/$ndf=1.9 for (a) and $\chi^2/$ndf=2.1 for (b).  
}
\label{fig:fit3param}
\end{figure}

%%%%%%%%%%%%%%%%%%%%%%%%%%%%%%%%%%%%%%%
\subsection{Jet masses for light-flavor jets}
%%%%%%%%%%%%%%%%%%%%%%%%%%%%%%%%%%%%%%%

The mass distribution of jets originating  from light quarks and gluons is distinct 
from jet masses initiated by top quarks. 
Figure~\ref{fig:fit3param}(a) shows the $\mjet$  distributions for light-flavor QCD jets
(without the $t\bar{t}$ process) for \pythia  and \herwig  after the fast detector simulation. 
The number of light-flavor jets was scaled to the expectation from the {\tt NLOjet++} program as discussed before.
Events with $W$+jet events where also studied in the
context of a possible contribution to the jet-mass shape. It was shown that $W$+jet events do not distort  
the region near $\mjet\simeq 170-180$ GeV. 

The jet masses for light jets can reasonably be described 
by the  functional form $a\cdot \mjet^{-b} \cdot \exp{(-c\cdot\mjet)}$,
where $a$, $b$ and $c$ are free parameters.  A similar function was previously used in the measurement of hadronic
$W/Z$ decays in two-jet mass spectra~\cite{Alitti:1990kw}.
A fit using this function provides a MC independent way to search for any significant deviations 
from the jet mass shape which is expected to be falling in the tails.  
The fit residuals for \pythia  and \herwig  show no significant deviation from zero.

\begin{figure}[ht] 
\begin{center}
\subfigure[Masses of jets with the $t\bar{t}$ signal scaled by two.]{
\includegraphics[scale=0.36, angle=0]{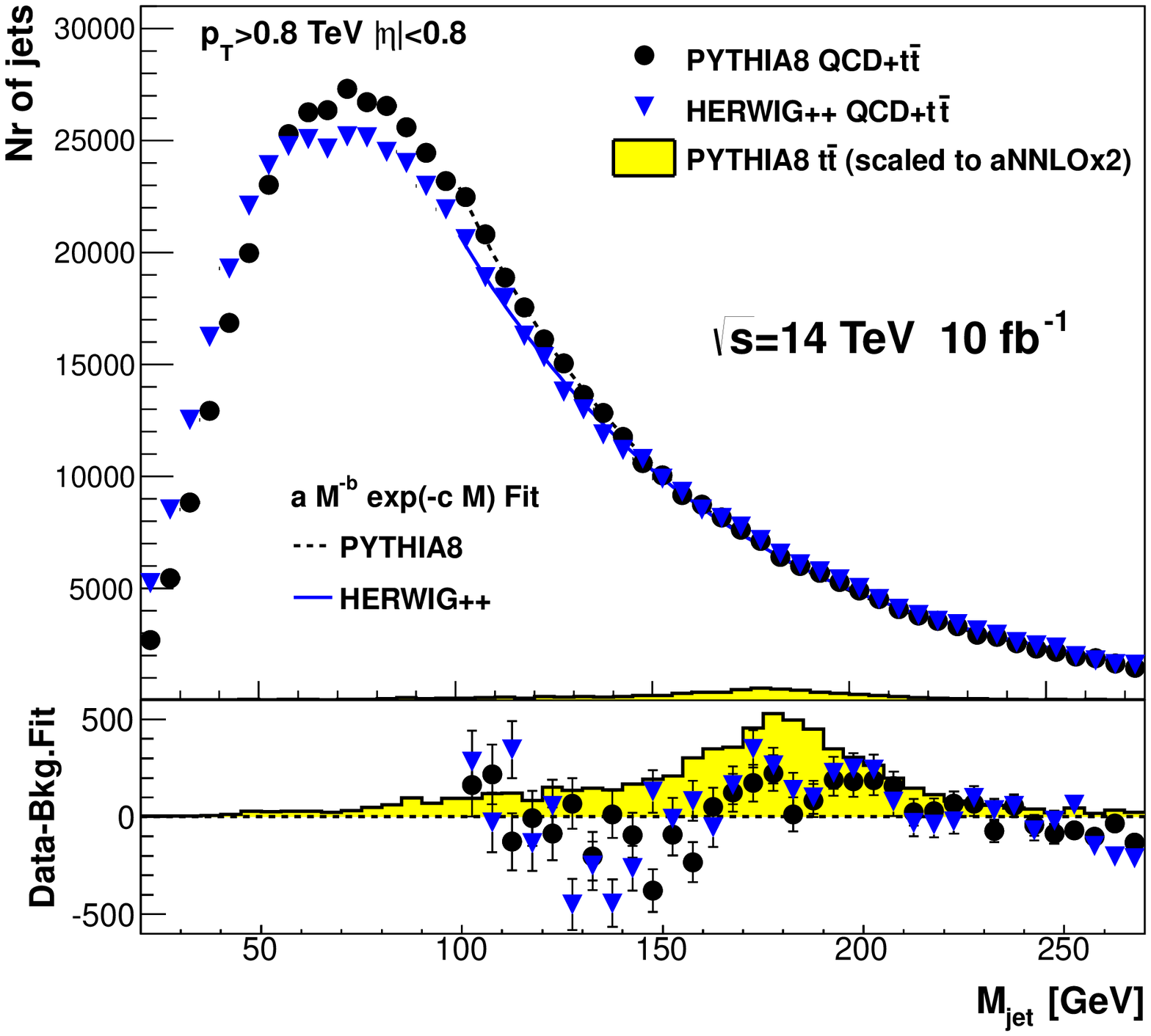}
}
\subfigure[Masses of jets with the $t\bar{t}$ signal after $b$-tagging.]{
\includegraphics[scale=0.36, angle=0]{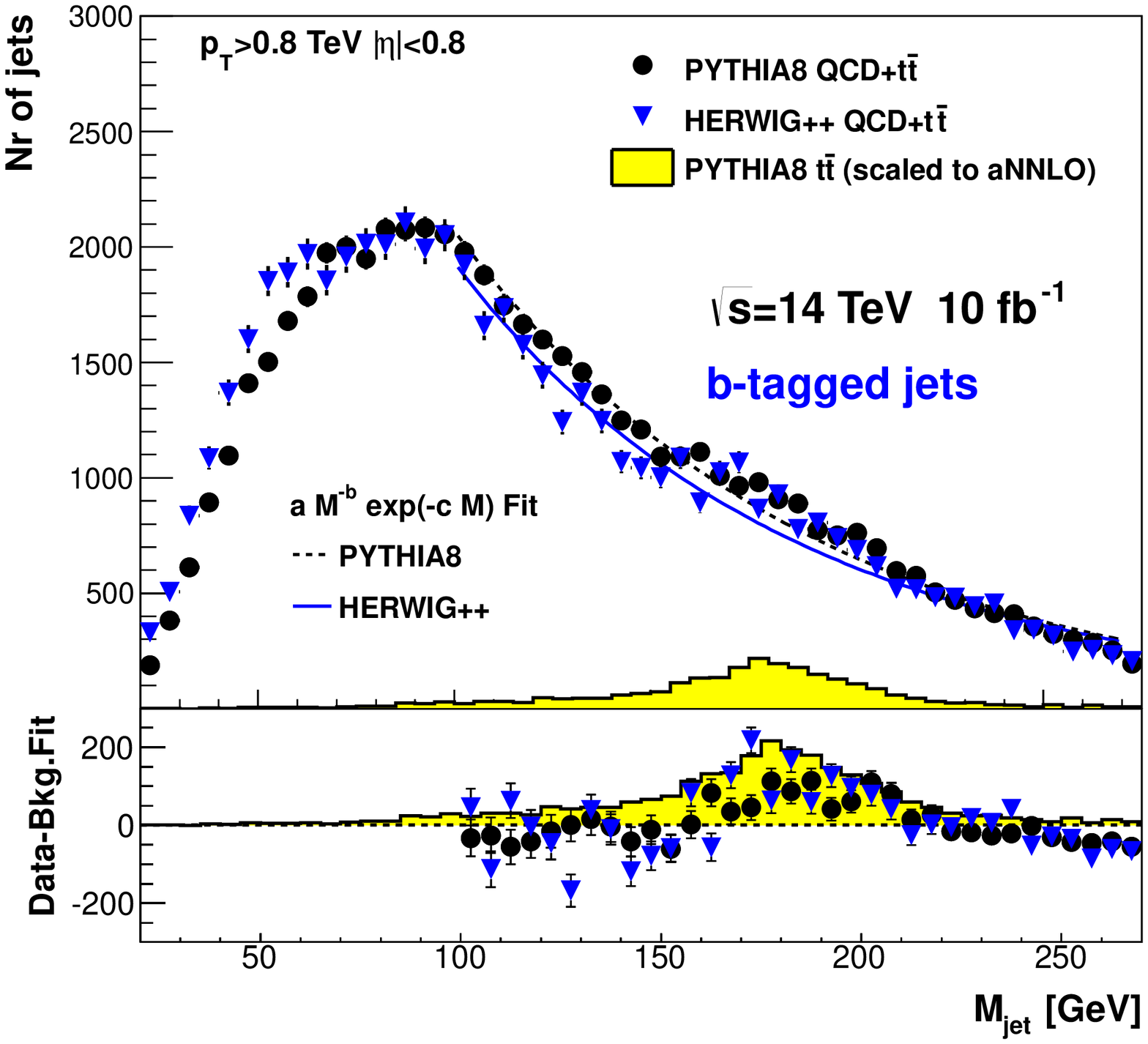}
}
\end{center}
\caption{
(a) Expectations for the  jet mass distributions using \pythia and \herwig after the
fast detector simulation. For the simulation,  top quarks were
added to light-flavor jets. The QCD dijet background
was scaled to the NLO inclusive jet cross section.
A $\chi^2$ fit was performed using the background function $a\cdot \mjet^{-b} \cdot \exp{(-c\cdot\mjet)}$
in the mass range $100<\mjet<270$~GeV.
(a) The \pythia expectation with the normalisation from
the aNNLO for $t\bar{t}$ was scaled by a factor two.
(b) The same distribution using the $t\bar{t}$ signal yield predicted by the aNNLO
after applying the $b$-tagging for background and top jets.
The fit quality using the background function is $\chi^2/$ndf=2.7 for (a) and $\chi^2/$ndf=3.5 for (b).
}
\label{fit3param_sig}
\end{figure}

\begin{figure}[ht]
\begin{center}
\subfigure[ {\tt PYTHIA} jet mass with the $t\bar{t}$ signal scaled by two.]{
\includegraphics[scale=0.36, angle=0]{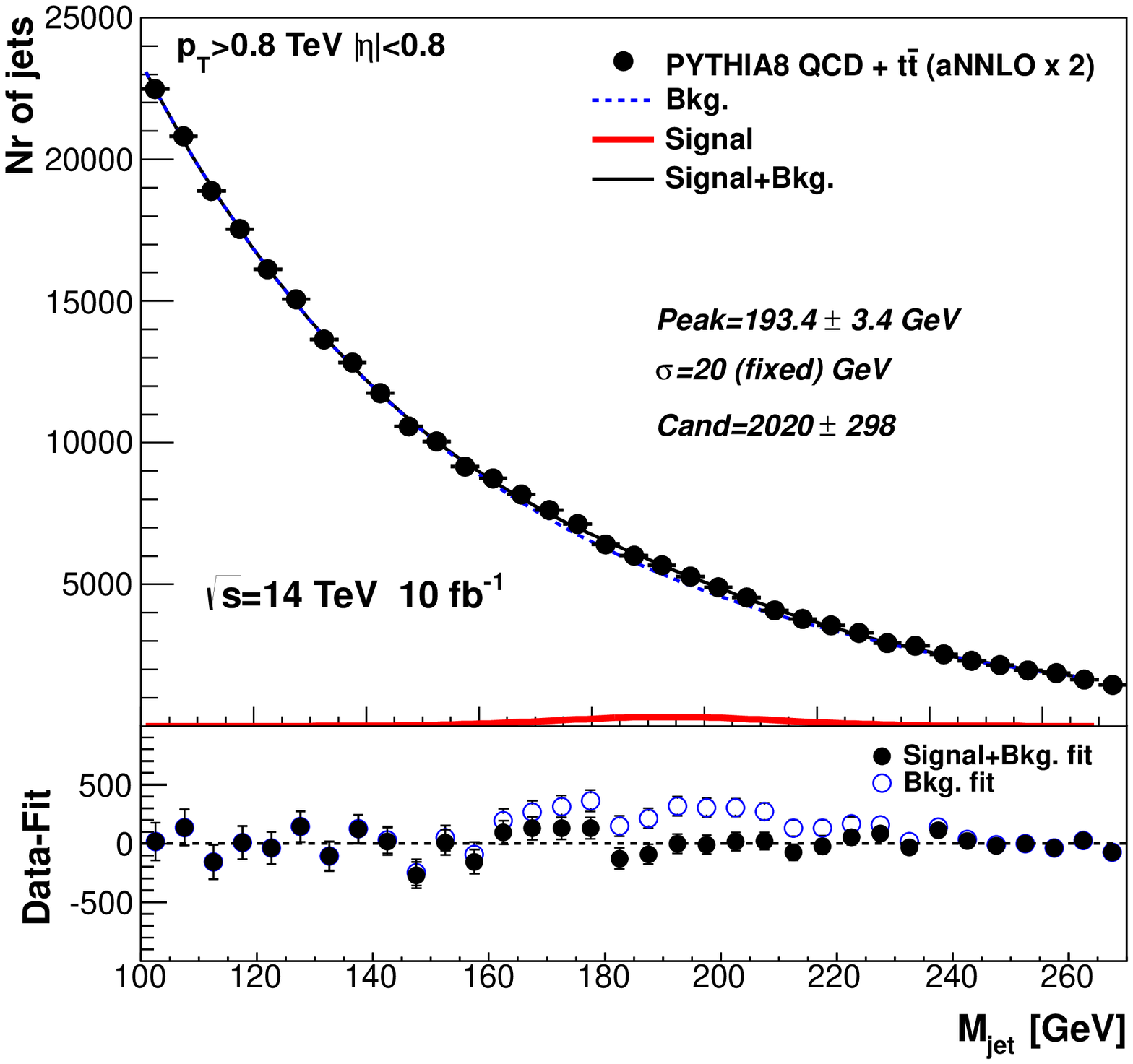}
}
\subfigure[ {\tt HERWIG} jet mass with the $t\bar{t}$ signal scaled by two.]{
\includegraphics[scale=0.36, angle=0]{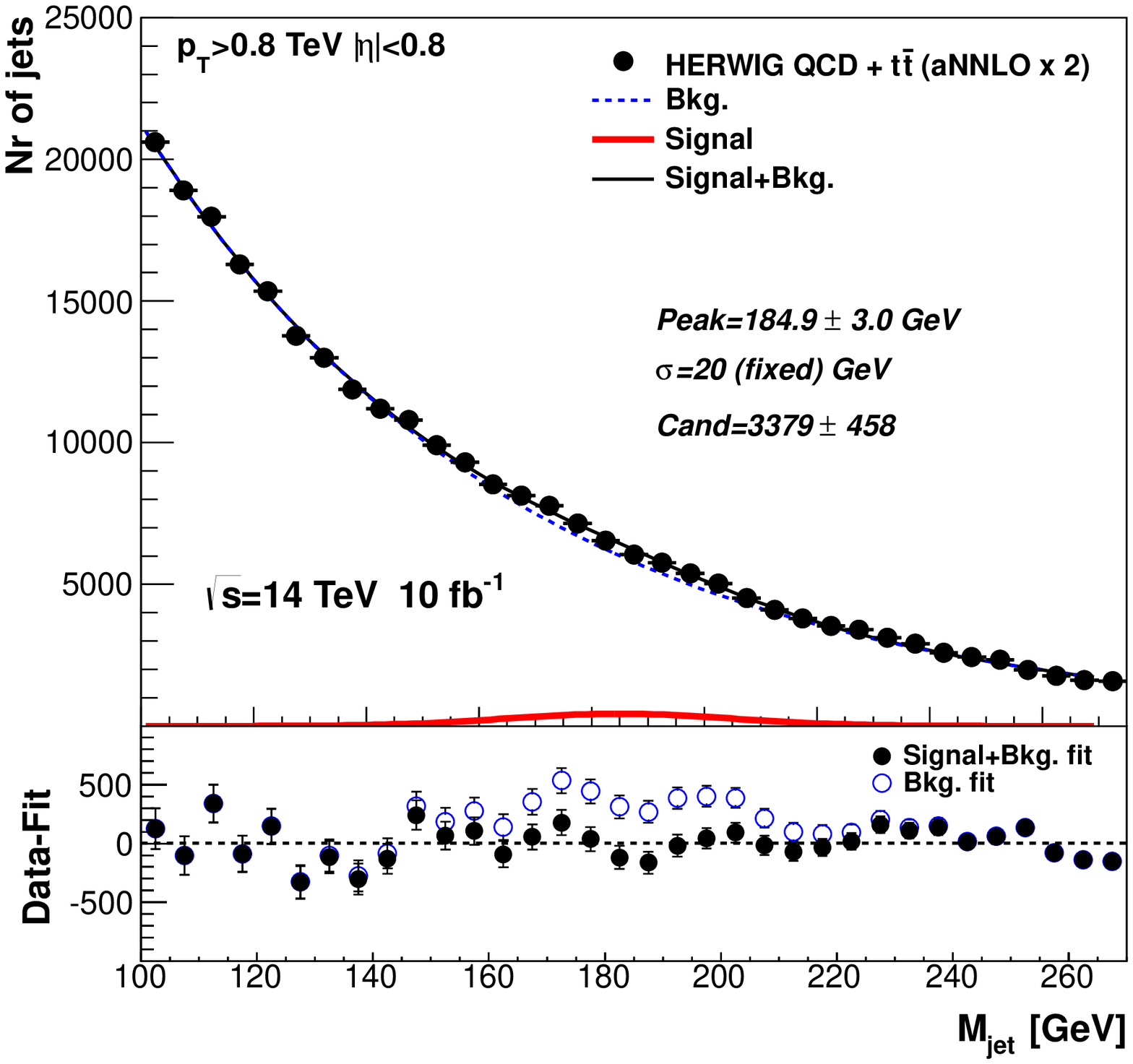}
}
\end{center}
\caption{
The distributions of jet masses  for $\ptjet >0.8$~TeV and $|\etajet|<0.8$ for MCs after
the fast detector simulation. The jet masses include contributions from $t\bar{t}$
assuming that the $t\bar{t}$ cross section is a factor two larger than the
aNNLO cross section.
A simultaneous $\chi^2$ fit was performed in the mass range $100<\mjet<270$~GeV using the
function $a\cdot \mjet^{-b} \cdot \exp{(-c\cdot\mjet)}$  for the
background description
plus a Gaussian to describe the excess near 170 GeV.
To improve the fit stability, the width of the Gaussian is fixed to 20~GeV as expected for top jets.
The bottom plots show the fit residuals with respect to the fitted signal
plus background function, as well as with respect to the
background component of the combined fit.
}
\label{fit_signalback}
\end{figure}

The inclusion of top quarks modifies $\mjet$  near the 170 GeV region. 
Figure~\ref{fig:fit3param}(b) shows the expectation for $\mjet$ assuming the contribution from 
the $t\bar{t}$ process. The top jets were simulated using \pythia, while their yield 
was scaled to the aNNLO calculation.
The fit using $a\cdot \mjet^{-b} \cdot \exp{(-c\cdot\mjet)}$ was performed in the range
$100<\mjet<270$~GeV. 
The fit residuals
do not show any significant excess above zero, indicating 
that the extraction of the top signal assuming the nominal aNNLO yield for 
$t\bar{t}$ can be challenging.

The situation is different if the top-quark yield is somewhat larger 
than the $t\bar{t}$ expectation. For example,
Fig.~\ref{fit3param_sig}(a) shows what happens when the top signal has a factor of two larger
cross section than the aNNLO prediction shown before. 
The signal is difficult to miss; the residuals of the fit near $\mjet\simeq 180$~GeV 
show an excess above zero
and have rather characteristic $S$-shape form due to the pull from the signal region. 
This is more apparent for \herwig 
than for \pythia, indicating a model dependence of this observation. 

Another way of looking at the effect of top quarks on the jet mass distribution is to reduce the contribution of 
light-flavor jets using a $b$-tagging technique. Figure~\ref{fit3param_sig}(b) shows the jet masses
with the nominal $t\bar{t}$ signal strength, but after applying a $b$-tagging using 
the {\tt DELPHES} \cite{Ovyn:2009tx} setting.  It assumes a 
$40\%$ $b-$quark reconstruction efficiency, $10\%$ and $1\%$ misstag rates due to $c-$quark and 
light-flavor jets, respectively. The $b$-tagging increases the signal-over-background ratio
and the $t\bar{t}$ signal is clearly observed.

The scenario when the cross section of boosted top quarks is higher than 
the  $t\bar{t}$ prediction was further studied in
Fig.~\ref{fit_signalback} when a potential excess of top jets near the 170 GeV region is 
extracted using a signal plus
background function. 
As before, light-flavor jets 
were combined with top jets from the $t\bar{t}$ process. 
For this hypothetical scenario, the  yield of the latter process 
was scaled by a factor of two with respect to the aNNLO prediction.
The signal function is assumed to be a Gaussian with the 
width of 20 GeV as expected for the top jets
(see Fig.~\ref{fig:tt}). 

The number of top quarks included in the simulation for $\pttop>0.65$~TeV was 
11,840 (5920 top quarks from the aNNLO times two, see Sect.~\ref{sec:theory}). 
This leads to 4,400  
top jets with $\ptjet>0.8$~TeV contributing to the $\mjet\simeq 170$ GeV region 
(2,200 top jets
in the Gaussian core shown in Fig.~\ref{fig:tt} times two).  
According to the fit shown in Fig.~\ref{fit_signalback}, the 
number of extracted top jets is between 2,000-3,400, depending on the MC simulation.
This number was extracted by integrating the Gaussian component of the background plus signal fit.
Thus, the extracted number of top jets is close to the expected number of 
top jets included in the simulation, but there is some indication
that the signal-plus-background fit somewhat 
underestimates the number of top jets.

While the scenario when the number of top jets 
is a factor of two  larger than the aNNLO 
prediction for $t\bar{t}$ may seem exotic at first,
such an assumption  may not be too far from the Standard Model expectation for top quarks 
produced inclusively within a jet (see the discussion in Sect.\ref{sec:theory}).
Given the large difference between the aNNLO and NLO \cite{Kidonakis:2010dk}, higher-order 
QCD effects for the $t\bar{t}$ process may play a significant role 
in an increase of top-quark jets at very large $\ptjet$. It is also important to mention that theoretical 
uncertainties, especially those related to PDF,
can be as large as $20\%$ (see Fig.~\ref{fig:nlo}). 
Less understood contributions from single-top production (about $30\%$ at lower $\pttop$),
flavor-changing processes and from fragmentation within jets  
should also be considered for the inclusive production of top quarks inside jets.
Taking into account all such effects, our conjecture about the factor two may not be too far
from the real situation.  
Therefore, a better understanding of all Standard Model processes 
leading to top production at  high $\ptjet$ is needed.

One can also consider the discussed  result from the point of view of discovery reach.
The approach can be used to exclude any potential source
of new physics for a number of models (such as those based on  $Z'$ and $W'$ bosons) 
leading to top  quarks at large $\ptjet$. From the above consideration, any source of new physics 
can be excluded if it leads to a top-quark cross section
above 1184~fb  in  the fiducial region $\ptjet>0.8$ TeV and $|\etajet|<0.8$.
This cross section
is obtained from the aNNLO $t\bar{t}$ prediction multiplied by a factor two.
Note that the approach  can exclude a number of exotic processes.     
For example, models with  $Z'$ and Kaluza–Klein gluons may have
larger cross sections compared to the Standard Model $t\bar{t}$ process at very large $\ptjet$. As a consequence,
a number of limits have been set \cite{Chatrchyan:2012cx,:2012qa} excluding such models 
up to $1.5-2$ TeV without experimental observation of top quarks from the Standard Model $t\bar{t}$ 
process at $\pttop>0.6$~TeV.

The high-precision studies of jet mass using analytic background templates  
may seem difficult from the instrumental point of view since we are looking for
a top quark signal on top of a smoothly falling distribution 
which has a signal-over-background ratio at the level of $10\%$ for the $t\bar{t}$ process.
However, the assessment of systematics on 
the presence of a bump must have a different strategy
than for a typical jet-mass measurement. Any variation of selection cuts 
or change in the instrumental procedure 
should be followed by the data-driven
approach using the analytic fit to identify a bump after each systematic change, unlike a typical QCD
measurement of jet masses. 
For example, jet-energy scale variation should lead to a change of jet masses,
but the signal strength after the signal-plus-background fit should not be strongly affected given 
the data-driven nature of such extraction.

Finally, a possibility of using other techniques based on $b$-tagging, jet shapes 
and jet substructure can be considered, which can also help to deal with some experimentally 
unavoidable effects, such as pile up. 
These techniques have the potential to increase the signal over background 
ratio for $\mjet$ close to 170 GeV when dealing with high-$p_T$ inclusive jets. 
This has been illustrated in Fig.~\ref{fit3param_sig}(b) when considering jets after the $b$-tagging. 
As follows from this study,
if the QCD multijet background is reduced 
at least by a factor of two compared to the top-quark signal,
the $t\bar{t}$ process should be well observed for the yield  
expected from the aNNLO calculation.
Studies of such techniques are outside the scope of this paper and can be found elsewhere
\cite{Agashe:2006hk,*Lillie:2007yh,*Butterworth:2007ke,*Almeida:2008tp,*Almeida:2008yp,
*Kaplan:2008ie,*Brooijmans:2008,*Butterworth:2009qa,*Ellis:2009su,*ATL-PHYS-PUB-2009-081,*CMS-PAS-JME-09-001,*Almeida:2010pa,*Hackstein:2010wk,Chekanov:2010vc,*Chekanov:2010gv}.

\section{Conclusions}

This paper shows that jet masses alone, 
without any complicated techniques involving substructure variables,
already provide a  sensitive probe for inclusively produced top quarks within high-$p_T$ jets.
Due to the nature of the inclusive measurement, 
such technique is not based on tagging of  top quarks in the opposite direction.
The approach allows  to study  top quarks using
the assumption that the background fit function has a  smoothly falling shape and 
does not contain a hump near the 170 GeV region, thus
it can be modeled analytically.

As shown in this paper,  the  method has the potential to  detect highly-boosted 
top quarks if their yield is a factor of two or more  larger than that 
from the best-understood $t\bar{t}$ process assuming the aNNLO prediction. 
This observation also implies that any technique capable of reducing  
QCD background near $\mjet\simeq 170$~GeV at least by a factor of two should be sufficient
for the observation of boosted top quarks from the Standard Model $t\bar{t}$ process.
There are other sources for inclusive production of top quarks  for very large $\ptjet$,
but their good understanding requires further studies.
Once they are understood, any enhancement of top-quark cross section over the Standard Model prediction
would be indicative of the presence of new resonances at the TeV scale.

\section*{Acknowledgements}
We would like to thank many colleagues for the discussion of these
results. We thank R.~Blair, T.~LeCompte, J.~Proudfoot and R.~Yoshida for the discussion 
of the jet-mass fitting technique.   We also thank M.~Schulze,  E.~Berger  and Z.~Sullivan 
for the discussion and their help with the NLO calculations.

The submitted manuscript has been created by UChicago Argonne, LLC,
Operator of Argonne National Laboratory (``Argonne'').
Argonne, a U.S. Department of Energy Office of Science laboratory,
is operated under Contract No. DE-AC02-06CH11357.
This research used resources of the Argonne Leadership 
Computing Facility at Argonne National Laboratory, which is supported by the 
Office of Science of the U.S. Department of Energy under contract DE-AC02-06CH11357.
The work of N. Kidonakis is supported by the National Science Foundation 
under Grant No. PHY 1212472.

%%%%%%%%%%%%%%%%%%%%%% references %%%%%%%%%%%%%%%%%%%%%%%%%%%%%%
% \bibliographystyle{./Macros/h-elsevier-clean}
% \bibliographystyle{./Macros/l4z_pl}
% \bibliographystyle{./Macros/l4z_epj.bst}
% \bibliographystyle{./Macros/l4z_np}
% \bibliographystyle{./Macros/elsart-num}
%  \def\bibname{\Large\bf References}
% \def\refname{\Large\bf References}
% \pagestyle{plain}
\bibliography{biblio}

\end{document}